\documentstyle[12pt,fleqn]{article}

\font\twlgot =eufm10 scaled \magstep1
\font\egtgot =eufm8
\font\sevgot =eufm7

\font\twlmsb =msbm10 scaled \magstep1
\font\egtmsb =msbm8
\font\sevmsb =msbm7

\newfam\gotfam
\def\pgot{\fam\gotfam\twlgot}
\textfont\gotfam\twlgot
\scriptfont\gotfam\egtgot
\scriptscriptfont\gotfam\sevgot
\def\got{\protect\pgot}

\newfam\msbfam
\textfont\msbfam\twlmsb
\scriptfont\msbfam\egtmsb
\scriptscriptfont\msbfam\sevmsb
\def\Bbb{\protect\pBbb}
\def\pBbb{\relax\ifmmode\expandafter\Bb\else\typeout{You cann't use
Bbb in text mode}\fi}
\def\Bb #1{{\fam\msbfam\relax#1}}

\newcommand{\gO}{{\got O}}
\newcommand{\gQ}{{\got T}}
\newcommand{\gU}{{\got U}}
\newcommand{\gE}{{\got E}}

\newcommand{\gP}{{\got P}}
\newcommand{\gX}{{\got X}}

\def\thebibliography#1{\section*{References}\list
  {[\arabic{enumi}]}{\settowidth\labelwidth{#1}\leftmargin\labelwidth
    \advance\leftmargin\labelsep
    \usecounter{enumi}}
    \def\newblock{\hskip .11em plus .33em minus .07em}
    \sloppy\clubpenalty4000\widowpenalty4000
    \sfcode`\.=1000\relax}

\let\Large=\large

\def\op#1{\mathop{\fam0 #1}\limits}

\newcommand{\im}{{\rm Im\,}}
\newcommand{\di}{{\rm dim\,}}

\newcommand{\Ker}{{\rm Ker\,}}
\newcommand{\nm}[1]{\mid {#1}\mid}

\newcommand{\beq}{\begin{equation}}
\newcommand{\eeq}{\end{equation}}
\newcommand{\ben}{\begin{eqnarray}}
\newcommand{\een}{\end{eqnarray}}
\newcommand{\be}{\begin{eqnarray*}}
\newcommand{\ee}{\end{eqnarray*}}
\newcommand{\bea}{\begin{eqalph}}
\newcommand{\eea}{\end{eqalph}}

\newcommand{\cP}{{\cal P}}

\newcommand{\cO}{{\cal O}}

\newcommand{\cQ}{{\cal T}}
\newcommand{\cE}{{\cal E}}

\newcommand{\bs}{{\bf s}}

\newcommand{\al}{\alpha}

\newcommand{\dl}{\delta}
\newcommand{\la}{\lambda}
\newcommand{\La}{\Lambda}
\newcommand{\f}{\phi}
\newcommand{\om}{\omega}

\newcommand{\m}{\mu}

\newcommand{\g}{\gamma}
\newcommand{\G}{\Gamma}
\newcommand{\th}{\theta}

\newcommand{\si}{\sigma}

\newcommand{\w}{\wedge}
\newcommand{\wt}{\widetilde}

\newcommand{\ol}{\overline}
\newcommand{\dr}{\partial}
\newcommand{\ar}{\op\longrightarrow}

\newcounter{eqalph}
\newcounter{equationa}
\newcounter{theorem}
\newcounter{proposition}
\newcounter{lemma}
\newcounter{corollary}
\newcounter{definition}

\def\thedefinition{\arabic{definition}}

\newenvironment{proof}{\noindent {\it Proof.}}{\hfill $\Box$
\medskip }

\newenvironment{theo}{\refstepcounter{definition} \medskip\noindent
THEOREM \thedefinition.\it}{\medskip }
\newenvironment{prop}{\refstepcounter{definition} \medskip\noindent
PROPOSITION \thedefinition.\it}{\medskip }
\newenvironment{lem}{\refstepcounter{definition} \medskip\noindent  LEMMA
\thedefinition.\it }{\medskip }
\newenvironment{cor}{\refstepcounter{definition} \medskip\noindent 
COROLLARY \thedefinition.\it }{\medskip }

\newenvironment{eqalph}{\stepcounter{equation}
\setcounter{equationa}{\value{equation}}
\setcounter{equation}{0}

\begin{eqnarray}}{\end{eqnarray}\setcounter{equation}{\value{equationa}}}

\newcommand{\mar}[1]{}

\begin{document}
\hbox{}

{\parindent=0pt 

{ \Large \bf Global calculus in BRST cohomology}
\bigskip

{\sc G. GIACHETTA, 
L. MANGIAROTTI$^1$  and
G.SARDANASHVILY$^2$}

{ \small

{\it $^1$ Department of Mathematics and Physics, University of Camerino,
62032 Camerino (MC), Italy, e-mail: mangiaro@camserv.unicam.it

$^2$ Department of Theoretical Physics,
Physics Faculty, Moscow State University, 117234 Moscow, Russia, e-mail:
sard@grav.phys.msu.su}
\bigskip

{\bf Abstract} 
The iterated BRST
cohomology is studied by computing
cohomology of the variational complex on the infinite order
jet space of a smooth fibre bundle.  This computation also provides a
solution of the global inverse problem of the calculus of variations in
Lagrangian field theory.
\medskip

\medskip


} }

\section{Introduction}

We address the Lagrangian
antifield BRST formalism  of \cite{barn,barn00,brandt,henn91}, phrased in
terms of exterior forms of finite jet order on
the infinite order jet space of physical fields, ghosts and antifields on a
base manifold $X$. Horizontal (semibasic) exterior forms 
constitute a bicomplex with respect to the BRST operator
$\bs$ and the horizontal (total) differential
$d_H$. It is graded by the ghost number $k$ and the form degree $p$.
We aim to study the iterated $\bs$-cohomology $H^{k,p}(\bs|d_H)$
of the
$d_H$-cohomology groups of this bicomplex (i.e., 
the term $E_2^{*,*}$ of its spectral sequence \cite{mcl}). 
In terms of form degree $p=n=\dim X$, this cohomology
coincides with the well-known local BRST cohomology (i.e., $\bs$-cohomology
modulo $d_H$). If $p<n$, iterated BRST
cohomology unlike local BRST cohomology is defined only for
$d_H$-closed forms.  

The above mentioned 
BRST formalism is usually
formulated on a contractible base
$X$ when the $d_H$-cohomology of form degree $0<p<n$ are trivial in
accordance with the algebraic Poincar\'e lemma (see Lemma \ref{am12} below).
If $X=\Bbb R^n$, there is an isomorphism of the iterated 
(and local) BRST cohomology groups $H^{k,n}(\bs|d_H)$, $k\neq -n$, to the
cohomology
$H^{k+n}_{\rm tot}$ groups of the total BRST operator
$\wt\bs=\bs +d_H$ on horizontal exterior forms of total ghost number $k+n$
\cite{brandt}. We generalize this result to the case of iterated BRST
cohomology of form degree $p<n$ and an arbitrary connected manifold $X$
(see Corollary \ref{aaa} below). To construct the corresponding (global) 
descent equations, one needs the $d_H$-cohomology of exterior forms on
the infinite order jet space. The study of this cohomology is the key
point of our consideration. 

Note that the descent equations for representativers of local BRST
cohomology groups of form degree $p<n$ are also constructed,
but this cohomology fails to be related to cohomology of the total
BRST operator \cite{barn00} (see also \cite{db92} where BRST cohomology
modulo the exterior differential $d$ in the Yang--Mills gauge theory are
considered). 

To avoid the
sophisticate techniques of ghosts and antifields \cite{book00}, we here study
iterated  cohomology of the graded differential algebra
$\cO^*_\infty$ of exterior forms of finite jet order on the infinite order jet
space $J^\infty Y$ of an affine bundle
$Y\to X$. 
Note that 
affine bundles provide a standard framework in quantum field theory.
Let $\cO^*_\infty$ be endowed with a nilpotent form degree preserving
endomorphism
$\bs$ such that  
horizontal elements of
$\cO^*_\infty$ constitute a bicomplex with respect to $\bs$ and $d_H$, i.e.,
$\bs d_H+ d_H\bs=0$. We agree to call a gradation degree $k$ with
respect to
$\bs$  the ghost number. Suppose that $\bs$  vanishes on exterior forms on $X$
and that these forms are of zero ghost number and are not $\bs$-exact.
The goal is the following. 

\begin{theo} \label{aa2} \mar{aa2}
If $Y\to X$ is an affine 
bundle, the iterated $\bs|d_H$-cohomology is the following. 

(i) The $H^{\neq 0, p<n}(\bs|d_H)$ is trivial. 

(ii) $H^{0,p<n}(\bs|d_H)= H^p(X)$ where $H^*(X)$ is de Rham cohomology of $X$.

(iii) $H^{k,n}(\bs|d_H)=H^{k+n}_{\rm
tot}$, $k<-n$ or $k> -1$.

(iv)  Let $\g_p
: H^p(X)\to H^p_{\rm tot}$, $0\leq p<n$, be a natural homomorphism
corresponding to the monomorphism of the algebra $\cO^*(X)$ of exterior forms
on $X$ to $\cO^*_\infty$. Put
$\ol H^p=H^p_{\rm tot}/\im \g_p$. If $-n\leq k< -1$, 
there is a monomorphism $\ol H^{k+n}\to H^{k,n}(\bs|d_H)$ such that
$H^{k,n}(\bs|d_H)/\ol H^{k+n}=\Ker \g_{k+n+1}$. 
In particular, $\Ker \g_0 =0$
and $\ol H^0=H^0_{\rm tot}/H^0(X)$. 

(v) $H^{-1,n}(\bs|d_H)= \ol H^{n-1}$.
\end{theo}

Note that the operator $\bs$ in Theorem \ref{aa2} may have different physical
origins. The following corollary of Theorem \ref{aa2} corresponds to the case
of iterated BRST cohomology.

\begin{cor} \label{aaa} \mar{aaa}
Let $Y\to X$ be a vector bundle and $P^*_\infty\subset \cO^*_\infty$ a
subalgebra of exterior forms which are polynomial in fibre coordinates of
$J^\infty Y\to X$. There is the decomposition of
$C^\infty(X)$-modules $P^*_\infty=\cO^*(X)\oplus (P^*_\infty)_{>0}$. Let
$\wt\bs(P^*_\infty)\subset (P^*_\infty)_{>0}$. Then, $\g_p$, $0\leq p<n$, are
monomorphisms,  and the items (iv), (v) of Theorem \ref{aa2} state 
isomorphisms $H^{k,n}(\bs|d_H)=H^{k+n}_{\rm tot}/H^{k+n}(X)$, 
$-n\leq k\leq -1$.
\end{cor}

In particular, if $X=\Bbb R^n$, the iterated BRST cohomology
of form degree $p<n$ (except $H^{0,0}(\bs|d_H)=\Bbb R$) is always
trivial in contrast with the local BRST cohomology.

In fact, to prove Theorem \ref{aa2}, we will obtain
$d_H$- and
$\dl$-cohomology of the variational complex of the graded differential algebra
$\cO^*_\infty$ in the case of an arbitrary smooth bundle $Y\to X$.

The $\cO^*_\infty$ is the direct
limit of graded differential algebras of exterior forms on finite
order jet manifolds. It consists of exterior
forms on finite order jet manifolds modulo the pull-back identification.
Passing to the direct limit of the de Rham complexes
of exterior forms on finite order jet manifolds, de Rham cohomology of 
$\cO^*_\infty$ has been found to coincide with de Rham cohomology
of the bundle $Y$
\cite{ander,bau}. However, this is not a way of studying other cohomology
groups of
$\cO^*_\infty$.  
Therefore, we enlarge 
$\cO^*_\infty$ to 
the structure algebra $\cQ^*_\infty$
of the sheaf of germs of exterior forms on finite order jet
manifolds. One can say that $\cQ^*_\infty$ consists of
exterior forms of locally finite jet order on $J^\infty Y$.  
The $d_H$- and $\dl$-cohomology of $\cQ^\infty$
has been investigated in \cite{tak2}. We simplify this investigation
due to Lemma \ref{lmp03} below and  prove that
$\cO^*_\infty$ and
$\cQ^*_\infty$ have the same $d_H$- and $\dl$-cohomology (see Theorem
\ref{am11} below). In particular, this provides a solution of the global
inverse problem of the calculus of variations in the class of exterior forms
of finite jet order. 

For the proof of Theorem \ref{aa2}, it is quite
important that, if $Y\to X$ is an affine bundle,
$d_H$-cohomology $H^{<n}(d_H;\cO^*_\infty)$ of $\cO^*_\infty$ coincides with de
Rham cohomology of the base $X$. It follows
that every $d_H$-closed $(k<n)$-form $\f\in\cO^*_\infty$ splits into the sum
$\f=\varphi + d_H\xi$ where $\xi\in \cO^*_\infty$
and $\varphi$ is a closed form on $X$. 
Since the operator $\bs$ annihilates these forms, the system of global descent
equations can be constructed though its right-hand side is not 
zero. 
We come to the same result for the polynomial algebra
$P^*_\infty$ 
and its subalgebra $\ol P^*_\infty$ of $x$-independent forms.

\section{The differential calculus on $J^\infty Y$} 

Smooth manifolds throughout are assumed to be
real, finite-dimensional, Hausdorff,
paracompact, and connected. Put further dim$X=n$. The standard
notation of jet formalism \cite{book,book00} is utilized. 
Following the terminology of
\cite{hir}, by a sheaf $S$ on a topological space $Z$ is meant a sheaf
bundle $S\to Z$. Accordingly, $\G(S)$ denotes the canonical presheaf of
sections of the sheaf $S$, and 
$\G(Z,S)$ is the set of global sections of $S$. 

Recall that the infinite order jet space $J^\infty Y$ of a smooth bundle
$Y\to X$ is defined as a projective limit
$(J^\infty Y,\pi^\infty_r)$ of the inverse system
\mar{t1}\beq
X\op\longleftarrow^\pi Y\op\longleftarrow^{\pi^1_0}\cdots \longleftarrow
J^{r-1}Y \op\longleftarrow^{\pi^r_{r-1}} J^rY\longleftarrow\cdots \label{t1}
\eeq
of finite order jet manifolds $J^rY$ of $Y\to X$, where $\pi^r_{r-1}$ are
affine bundles. Bearing in mind Borel's theorem, one can say that 
$J^\infty Y$ consists of the equivalence classes of sections of $Y\to X$
identified by their Taylor series at points of $X$.
Endowed with the projective limit topology,
$J^\infty Y$ is a paracompact Fr\'echet manifold \cite{tak2}. A bundle
coordinate atlas
$\{U_Y,(x^\la,y^i)\}$ of $Y$ yields the manifold
coordinate atlas $\{(\pi^\infty_0)^{-1}(U_Y), (x^\la, y^i_\La)\}$, 
$0\leq|\La|$, of $J^\infty
Y$, together with the transition functions  
\mar{55.21}\beq
{y'}^i_{\la+\La}=\frac{\dr x^\m}{\dr x'^\la}d_\m y'^i_\La, \label{55.21}
\eeq
where $d_\mu$ denotes the total derivative 
$d_\mu = \dr_\mu + \op\sum_{|\La|\geq 0} y^i_{\mu+\La}\dr_i^\La$.

With the inverse system (\ref{t1}),
one has the direct system 
\mar{t20}\beq
\cO^*(X)\op\longrightarrow^{\pi^*} \cO^*_0 
\op\longrightarrow^{\pi^1_0{}^*} \cO_1^*
\op\longrightarrow^{\pi^2_1{}^*} \cdots \op\longrightarrow^{\pi^r_{r-1}{}^*}
 \cO_r^* \longrightarrow\cdots \label{t20}
\eeq
of $\Bbb R$-algebras $\cO^*_r$ of exterior forms on finite
order jet manifolds, where $\pi^r_{r-1}{}^*$ are pull-back
monomorphisms. Its direct limit is the above mentioned  graded
differential $\Bbb R$-algebra $\cO^*_\infty$.
It is a
differential calculus over the $\Bbb R$-ring $\cO^0_\infty$ of continuous real
functions on
$J^\infty Y$ which are the pull-back of smooth functions on
finite order jet manifolds. 
Let us enlarge the ring $\cO^0_\infty$ to the $\Bbb
R$-ring
$\cQ^0_\infty$ of continuous real functions on $J^\infty Y$ such that, given
$f\in
\cQ^0_\infty$ and any point $q\in J^\infty Y$, there exists a neighborhood
of $q$ where $f$ coincides with the pull-back of a smooth function on some
finite order jet manifold.  Let $\gO^*_r$ be a sheaf
of germs of exterior forms on the $r$-order jet manifold $J^rY$ and 
$\G(\gO^*_r)$ its canonical presheaf.  There is the direct system of canonical
presheaves
\be
\G(\gO^*_X)\op\longrightarrow^{\pi^*} \G(\gO^*_0) 
\op\longrightarrow^{\pi^1_0{}^*} \G(\gO_1^*)
\op\longrightarrow^{\pi^2_1{}^*} \cdots \op\longrightarrow^{\pi^r_{r-1}{}^*}
 \G(\gO_r^*) \longrightarrow\cdots. 
\ee
 Its direct
limit $\gO^*_\infty$ 
is a presheaf of graded differential
$\Bbb R$-algebras on
$J^\infty Y$. Let $\gQ^*_\infty$ be the sheaf constructed from 
$\gO^*_\infty$ and $\G(\gQ^*_\infty)$ its canonical presheaf. 
The structure algebra 
$\cQ^*_\infty=\G(J^\infty Y,\gQ^*_\infty)$ of
the sheaf $\gQ^*_\infty$  
is a differential calculus over the $\Bbb
R$-ring $\cQ^*_\infty$. 
There are the 
$\Bbb R$-algebra monomorphisms 
 $\gO^*_\infty \to\G(\gQ^*_\infty)$ and $\cO^*_\infty \to\G(\cQ^*_\infty)$. 
Since the paracompact space
$J^\infty Y$ admits a partition of unity by elements of
$\cQ^0_\infty$ \cite{tak2}, sheaves of
$\cQ^0_\infty$-modules on
$J^\infty Y$ are fine and acyclic. Then, the 
abstract de Rham theorem on cohomology of a sheaf resolution \cite{hir} can be
called into play in order to obtain cohomology of the algebra $\cQ^*_\infty$. 

For short, we
agree to call elements of $\cQ^*_\infty$ the
exterior forms on
$J^\infty Y$.  Restricted to a
coordinate chart
$(\pi^\infty_0)^{-1}(U_Y)$ of $J^\infty Y$, they
can be written in a coordinate form, where horizontal and contact forms
$\{dx^\la;\th^i_\La=dy^i_\La -y^i_{\la+\La}dx^\la\}$ provide
generators of the algebra
$\cQ^*_\infty$. 
There is the canonical decomposition of $\cQ^*_\infty$ into 
$\cQ^0_\infty$-modules $\cQ^{k,s}_\infty$
of $k$-contact and $s$-horizontal forms:
\be
\cQ^*_\infty =\op\oplus_{k,s}\cQ^{k,s}_\infty,
\qquad h_k:\cQ^*_\infty\to \cQ^{k,*}_\infty,  \qquad
h^s:\cQ^*_\infty\to \cQ^{*,s}_\infty,
 \quad 0\leq k,
\quad 0\leq s\leq n.
\ee
Accordingly, the
exterior differential on $\cQ_\infty^*$ 
splits into the sum $d=d_H+d_V$ of horizontal and vertical differentials
such that
\be
&& d_H\circ h_k=h_k\circ d\circ h_k, \qquad d_H(\f)=
dx^\la\w d_\la(\f), \qquad \f\in\cQ^*_\infty,\\ 
&& d_V\circ h^s=h^s\circ d\circ h^s, \qquad
d_V(\f)=\th^i_\La \w \dr^\La_i\f.
\ee

\section{The horizontal complex}

Being nilpotent, the
differentials $d_V$ and $d_H$ provide the natural bicomplex
$\{\cQ^{k,m}_\infty\}$ of  the graded differential algebra
$\cQ^*_\infty$. Let us consider its row
\mar{t70}\beq
0\to\Bbb R\to \cQ^0_\infty \ar^{d_H}\cQ^{0,1}_\infty\ar^{d_H}\cdots  
\op\longrightarrow^{d_H} 
\cQ^{0,n}_\infty\ar^{d_H} 0  \label{t70}
\eeq
called the horizontal complex. 
The corresponding complex of sheaves 
\mar{t70a}\beq
0\to\Bbb R\to \gQ^0_\infty \ar^{d_H}\gQ^{0,1}_\infty\ar^{d_H}\cdots  
\op\longrightarrow^{d_H} 
\gQ^{0,n}_\infty\ar^{d_H} 0  \label{t70a}
\eeq
except the last term is exact (see Lemma \ref{am12} below).
 Then, since
the sheaves $\gQ^{0,<n}$ of $\cQ^0_\infty$-modules on $J^\infty Y$ are fine,
we obtain from the abstract de Rham theorem and Lemma (\ref{20jpa}) below
that $d_H$-cohomology $H^{<n}(d_H;\cQ^*_\infty)$ of the horizontal complex
(\ref{t70}) is equal to de Rham cohomology $H^*(Y)$ of the bundle $Y$.
Theorem \ref{am11} below shows that $d_H$-cohomology 
$H^{<n}(d_H;\cO^*_\infty)$ of the horizontal complex
\mar{t70'}\beq
0\to\Bbb R\to \cO^0_\infty \ar^{d_H}\cO^{0,1}_\infty\ar^{d_H}\cdots  
\op\longrightarrow^{d_H} 
\cO^{0,n}_\infty\ar^{d_H} 0  \label{t70'}
\eeq
of the algebra $\cO^*_\infty$ is the same.
However, one should complete
the horizontal complex (\ref{t70}) in the variational one in order to say
something on the $n$th cohomology group of $d_H$ (see the relation (\ref{1j})
below).

\section{The variational complex}

Let us consider the variational operator 
$\dl=\tau\circ d$ on $\cQ^{*,n}_\infty$ where 
\be
\tau=\op\sum_{k>0}\frac1k\ol\tau\circ h_k\circ h^n, \quad
\ol\tau(\f)
=\op\sum_{0\leq\nm\La}(-1)^{\nm\La}\th^i\w [d_\La(\dr^\La_i\rfloor\f)], 
\quad \f\in \cQ^{>0,n}_\infty,
\ee
is 
the projection
$\Bbb R$-module endomorphism 
of $\cQ^*_\infty$ such that $\tau\circ d_H=0$ (see, e.g.,
\cite{bau,book,tul}). The $\dl$ is nilpotent, and obeys the
relation 
\mar{am13}\beq
\dl\circ\tau-\tau\circ d=0. \label{am13}
\eeq
Put
$\gE_k=\tau(\gQ^{k,n}_\infty)$, $E_k=\tau(\cQ^{k,n}_\infty)$, $k>0$.
Since
$\tau$ is a projector, there are isomorphisms 
\be
\G(\gE_k)=\tau(\G(\gQ^{k,n}_\infty)), \qquad E_k=\G(J^\infty Y,\gE_k).
\ee
With operators 
$d_H$ and $\dl$, we have the variational complex 
\mar{tams1}\beq
0\to\Bbb R\to \gQ^0_\infty \ar^{d_H}\gQ^{0,1}_\infty\ar^{d_H}\cdots  
\op\longrightarrow^{d_H} 
\gQ^{0,n}_\infty  \op\longrightarrow^\dl \gE_1 
\op\longrightarrow^\dl 
\gE_2 \longrightarrow \cdots  \label{tams1}
\eeq
of the sheaf $\gQ^*_\infty$ and that 
\mar{b317}\beq
0\to\Bbb R\to \cQ^0_\infty \ar^{d_H}\cQ^{0,1}_\infty\ar^{d_H}\cdots  
\op\longrightarrow^{d_H} 
\cQ^{0,n}_\infty  \op\longrightarrow^\dl E_1 
\op\longrightarrow^\dl 
E_2 \longrightarrow \cdots  \label{b317}
\eeq
of its
structure algebra $\cQ^*_\infty$. The similar variational complex
$\{\cO^*_\infty, \ol E_k\}$ of the algebra
$\cO^*_\infty$ takes place. There are
the well-known statements summarized usually as the algebraic Poincar\'e lemma
(see, e.g.,
\cite{olver,tul}). 

\begin{lem} \label{am12} \mar{am12}
If $Y$ is a contractible bundle $\Bbb R^{n+p}\to\Bbb R^n$, the
variational complex $\{\cO^*_\infty, \ol E_k\}$ of the graded differential
algebra
$\cO^*_\infty$ is exact.
\end{lem}

It follows that the variational
complex of sheaves (\ref{tams1}) is exact for any smooth bundle $Y\to X$.
Moreover, the sheaves
$\gQ^{*,n}_\infty$ in this complex are fine, and so are the sheaves $\gE_k$ in
accordance with Lemma \ref{lmp03} below.
Hence, the variational complex (\ref{tams1}) is a resolution of the
constant sheaf $\Bbb R$ on $J^\infty Y$. 

\begin{lem} \label{lmp03} \mar{lmp03}
Sheaves $\gE_k$, $k>0$, are fine.
\end{lem}

\begin{proof}
Though the $\Bbb R$-modules $E_{k>1}$ fail to be
$\cQ^0_\infty$-modules \cite{tul}, one can use the fact that the sheaves
$\gE_{k>0}$ are projections $\tau(\gQ^{k,n}_\infty)$ of sheaves of
$\cQ^0_\infty$-modules. Let $\gU =\{U_i\}_{i\in I}$
be a 
locally finite open covering  of
$J^\infty Y$ and $\{f_i\in\cQ^0_\infty\}$ the associated partition of unity. 
For any open subset $U\subset J^\infty Y$ and any section
$\varphi$ of 
the sheaf $\gQ^{k,n}_\infty$ over $U$, let us put
$h_i(\varphi)=f_i\varphi$.
The endomorphisms $h_i$ of $\gQ^{k,n}_\infty$ yield the $\Bbb R$-module
endomorphisms 
\be
\ol h_i=\tau\circ h_i: \gE_k\ar^{\rm in} \gQ^{k,n}_\infty \ar^{h_i}
\gQ^{k,n}_\infty \ar^\tau \gE_k
\ee
of the sheaves $\gE_k$.
They possess the properties
required for $\gE_k$ to be a fine sheaf. Indeed, for each $i\in I$, ${\rm
supp}\,f_i\subset U_i$ provides a closed set  such that $\ol h_i$ is zero
outside this set, while the sum $\op\sum_{i\in I}\ol h_i$ is the identity
morphism.
\end{proof}

\section{Cohomology of $\cQ^*_\infty$}

\begin{lem} \label{20jpa} \mar{20jpa}
There is an
isomorphism 
\mar{lmp80}\beq
H^*(J^\infty Y,\Bbb R)= H^*(Y,\Bbb R)=H^*(Y) \label{lmp80}
\eeq
between cohomology $H^*(J^\infty Y,\Bbb R)$ of $J^\infty Y$ with
coefficients in the constant sheaf $\Bbb R$, that $H^*(Y,\Bbb R)$ of $Y$, and
de Rham cohomology $H^*(Y)$ of $Y$. 
\end{lem}

\begin{proof}
Since $Y$ is a strong deformation retract of $J^\infty Y$ (see, e.g.,
\cite{eprint}), the first isomorphism in (\ref{lmp80}) follows from the
Vietoris--Begle theorem
\cite{bred}, while the second
one results from the well-known de Rham theorem.
\end{proof}

Let us consider the de Rham complex of sheaves 
\mar{lmp71} \beq
0\to \Bbb R\to
\gQ^0_\infty\op\longrightarrow^d\gQ^1_\infty\op\longrightarrow^d
\cdots
\label{lmp71}
 \eeq
on $J^\infty Y$ and the corresponding de Rham complex of their structure
algebras
\mar{5.13'} \beq
0\to \Bbb R\to
\cQ^0_\infty\op\longrightarrow^d\cQ^1_\infty\op\longrightarrow^d
\cdots\, .
\label{5.13'}
\eeq
The complex (\ref{lmp71}) is exact due to
the Poincar\'e lemma, and is a resolution of the constant sheaf $\Bbb R$ on
$J^\infty Y$ since sheaves $\gQ^r_\infty$ are fine. Then, the abstract de
Rham theorem and Lemma
\ref{20jpa} lead to the following.

\begin{prop} \label{38jp} \mar{38jp}
De Rham cohomology $H^*(\cQ^*_\infty)$ 
of the graded differential algebra
$\cQ^*_\infty$  is isomorphic to that $H^*(Y)$ of the bundle $Y$.
\end{prop}

It follows that every closed form $\f\in \cQ^*_\infty$
splits into the sum
\mar{tams2}\beq
\f=\varphi +d\xi, \qquad \xi\in \cQ^*_\infty, \label{tams2} 
\eeq
where $\varphi$ is a closed form on the bundle $Y$. 

Similarly, from the abstract de Rham theorem and Lemma 
\ref{20jpa},  we obtain the following. 

\begin{prop} \label{lmp05} \mar{lmp05}
There is an isomorphism
between $d_H$- and $\dl$-cohomology of the
variational complex (\ref{b317}) and de Rham cohomology of the bundle
$Y$, namely,
\be
H^{k<n}(d_H;\cQ^*_\infty)=H^{k<n}(Y), \qquad H^{k-n}(\dl;
\cQ^*_\infty)=H^{k\geq n}(Y).
\ee
\end{prop}

This isomorphism  recovers the results of \cite{ander80,tak2}, but
note also the following.
The relation (\ref{am13}) for $\tau$ and
the relation $h_0d=d_Hh_0$ for $h_0$ define  a homomorphisms of the
de Rham complex (\ref{5.13'}) of the algebra $\cQ^*_\infty$ to its variational
complex (\ref{b317}). The corresponding homomorphism of their cohomology
groups is an isomorphism by virtue of Proposition \ref{38jp} and Proposition
\ref{lmp05}. Then, the splitting (\ref{tams2}) leads to the following
decompositions.

\begin{prop} \label{t41} \mar{t41}
Any $d_H$-closed form $\si\in\cQ^{0,m}$, $m< n$, is represented by the sum
\mar{t60}\beq
\si=h_0\varphi+ d_H \xi, \qquad \xi\in \cQ^{m-1}_\infty, \label{t60}
\eeq
where $\varphi$ is a closed $m$-form on $Y$.
Any $\dl$-closed form $\si\in\cQ^{k,n}$, $k\geq 0$, splits into
\mar{t42}\ben
&& \si=h_0\varphi + d_H\xi, \qquad k=0, \qquad \xi\in \cQ^{0,n-1}_\infty,
\label{t42a}\\ 
&& \si=\tau(\varphi) +\dl(\xi), \qquad k=1, \qquad \xi\in \cQ^{0,n}_\infty,
\label{t42b}\\
&& \si=\tau(\varphi) +\dl(\xi), \qquad k>1, \qquad \xi\in E_{k-1},
\label{t42c}
\een
where $\varphi$ is a closed $(n+k)$-form on $Y$. 
\end{prop}
 
\section{Cohomology of $\cO^*_\infty$}

\begin{theo} \label{am11} \mar{am11}
Graded differential algebra $\cO^*_\infty$ has the same $d_H$- and
$\dl$-cohomology of the variational complex as $\cQ^*_\infty$.
\end{theo}

\begin{proof}
Let the common symbol $D$ stand for $d_H$ and
$\dl$. Bearing in mind decompositions
(\ref{t60}) -- (\ref{t42c}), it suffices to show that, if an
element
$\f\in
\cO^*_\infty$ is
$D$-exact in the algebra $\cQ^*_\infty$, then it is
so in the algebra
$\cO^*_\infty$.
Lemma \ref{am12} states that,
if
$Y$ is a contractible bundle and a $D$-exact form $\f$ on $J^\infty Y$
is of finite jet order
$[\f]$ (i.e., $\f\in\cO^*_\infty$), there exists an exterior form $\varphi\in
\cO^*_\infty$ on $J^\infty Y$ such that $\f=D\varphi$. Moreover, a glance at
the homotopy operators for $d_H$ and $\dl$ \cite{olver} shows that  the
jet order
$[\varphi]$ of $\varphi$ is bounded for all exterior forms $\f$ of fixed
jet order. Let us call this fact the finite exactness of the operator
$D$. Given an arbitrary bundle
$Y$, the finite exactness takes place on $J^\infty Y|_U$ over any open subset
$U$ of
$Y$ which is homeomorphic to a convex open subset of $\Bbb R^{\di Y}$.
Let us prove the following.

(i) Suppose that the finite exactness of the operator $D$ takes place on
$J^\infty Y$ over open subsets
$U$, $V$ of $Y$ and their non-empty overlap $U\cap V$. Then, it is also true on
$J^\infty Y|_{U\cup V}$.

(ii) Given a family $\{U_\al\}$ of disjoint open subsets of $Y$, let us
suppose that the finite exactness takes place on $J^\infty Y|_{U_\al}$ over
every subset $U_\al$ from this family. Then, it is true on $J^\infty Y$ over
the union
$\op\cup_\al U_\al$ of these subsets.

\noindent
If these assertions hold, the finite
exactness of
$D$ on $J^\infty Y$ takes place  because one can
construct the corresponding covering of the manifold $Y$
(\cite{bred2}, Lemma 9.5). 

\noindent
{\it Proof of (i)}. Let
$\f=D\varphi\in\cO^*_\infty$ be a $D$-exact form on
$J^\infty Y$. By assumption, it can be brought into the form
$D\varphi_U$ on $(\pi^\infty_0)^{-1}(U)$ and $D\varphi_V$ on
$(\pi^\infty_0)^{-1}(V)$, where
$\varphi_U$ and $\varphi_V$ are exterior forms of finite jet
order. Let us consider their difference $\varphi_U-\varphi_V$ on 
$(\pi^\infty_0)^{-1}(U\cap V)$. It is a $D$-exact form of finite jet
order which, by assumption, can be written as 
$\varphi_U-\varphi_V=D\si$ where 
$\si$ is also of finite jet order. 
Lemma
\ref{am20} below shows that $\si=\si_U +\si_V$ where
$\si_U$ and
$\si_V$ are exterior forms of finite jet order on $(\pi^\infty_0)^{-1}(U)$ and
$(\pi^\infty_0)^{-1}(V)$, respectively. Then, putting
\be
\varphi'|_U=\varphi_U-D\si_U, \qquad  
\varphi'|_V=\varphi_V+ D\si_V,
\ee
we have $\f =D\varphi'$ on $(\pi^\infty_0)^{-1}(U\cap V)$ where 
$\varphi'$ 
is of finite jet order. 

\noindent
{\it Proof of (ii)}. Let
$\f\in\cO^*_\infty$ be a $D$-exact form on
$J^\infty Y$.
The finite exactness on $(\pi^\infty_0)^{-1}(\cup
U_\al)$ holds since $\f=D\varphi_\al$ on every $(\pi^\infty_0)^{-1}(U_\al)$
and, as was mentioned above, the jet order
$[\varphi_\al]$ is bounded on the set of exterior forms
$D\varphi_\al$ of fixed jet order $[\f]$. 
\end{proof}

\begin{lem} \label{am20} \mar{am20}
Let $U$ and $V$ be open subsets of a bundle $Y$ and $\si\in
\gO^*_\infty$ an exterior form of finite jet order on
$(\pi^\infty_0)^{-1}(U\cap V)\subset J^\infty Y$. Then, $\si$ splits
into  a sum $\si_U+ \si_V$ of exterior forms $\si_U$ and $\si_V$ of finite jet
order on
$(\pi^\infty_0)^{-1}(U)$ and $(\pi^\infty_0)^{-1}(V)$, respectively. 
\end{lem} 

\begin{proof}
By taking a smooth partition of unity on $U\cup V$ subordinate to the cover
$\{U,V\}$ and passing to the function with support in $V$, one gets a
smooth real function
$f$ on
$U\cup V$ which is 0 on a neighborhood of $U-V$ and 1 on a neighborhood of
$V-U$ in $U\cup V$. Let $(\pi^\infty_0)^*f$ be the pull-back of $f$ onto
$(\pi^\infty_0)^{-1}(U\cup V)$. The exterior form $((\pi^\infty_0)^*f)\si$ is
zero on a neighborhood of $(\pi^\infty_0)^{-1}(U)$ and, therefore, can be
extended by 0 to $(\pi^\infty_0)^{-1}(U)$. Let us denote it $\si_U$.
Accordingly, the exterior form
$(1-(\pi^\infty_0)^*f)\si$ has an extension $\si_V$ by 0 to 
$(\pi^\infty_0)^{-1}(V)$. Then, $\si=\si_U +\si_V$ is a desired decomposition
because $\si_U$ and $\si_V$
are of finite jet order which does not exceed that of $\si$. 
\end{proof}

\section{The global inverse problem}

The expressions (\ref{t42a}) -- (\ref{t42b}) in Proposition \ref{t41} provide a
solution of the global inverse problem of the calculus of variations on fibre
bundles in the class of Lagrangians $L\in\cQ^{0,n}_\infty$ of locally finite
order \cite{ander80,tak2} (which is not so interesting for physical
applications). These expressions together with Theorem \ref{am11} give a
solution of the global inverse problem of the finite order calculus of
variations.

\begin{prop} \label{lmp112'} \mar{lmp112'}
(i) A finite order Lagrangian $L\in \cO^{0,n}_\infty$ is variationally trivial,
i.e.,  $\dl(L)=0$ iff 
\mar{tams3}\beq
L=h_0\varphi + d_H \xi, \qquad \xi\in \cO^{0,n-1}_\infty, \label{tams3}
\eeq
where $\varphi$ is a closed $n$-form on $Y$.
(ii) A finite order 
Euler--Lagrange-type operator satisfies the Helmholtz
condition $\dl(\cE)=0$ iff 
\be
\cE=\dl(L) + \tau(\f), \qquad L\in\cO^{0,n}_\infty, 
\ee
where $\f$ is a closed $(n+1)$-form on $Y$ (see also \cite{vin}).
\end{prop}

A solution of the global inverse problem of the fixed order calculus of
variations has been
suggested in
\cite{ander80} by computing cohomology of the fixed
order variational sequence.
However, the proof of the local exactness
of this variational sequence  requires
rather sophisticated {\it ad hoc} techniques in order to be reproduced 
(see also 
\cite{kru98}). The first thesis of \cite{ander80} agrees with that of
Proposition 
\ref{lmp112'}i, but says that the
jet order of the form $\xi$ in the expression (\ref{tams3}) is $k-1$ if $L$ is
a $k$-order variationally trivial Lagrangian. The second one states that a
$2k$-order Euler--Lagrange operator can be always  associated with a $k$-order
Lagrangian. 

One obtains from Proposition \ref{lmp112'}(i)  that 
the cohomology group
$H^n(d_H;\cO^*_\infty)$ of the complex (\ref{t70a}) obeys the relation
\mar{1j}\beq
H^n(d_H;\cO^*_\infty)/H^n(Y)=\dl(\cO^{0,n}_\infty), \label{1j}
\eeq 
where $\dl(\cO^{0,n}_\infty)$ is the $\Bbb R$-module of Euler--Lagrange
forms on $J^\infty Y$.

\section{The case of an affine bundle}

Let $Y\to X$ be an affine bundle. Since $X$ is a strong deformation retract
of $Y$, de Rham cohomology of $Y$ is equal to
that of
$X$. It leads to the cohomology
isomorphisms
\be
H^{<n}(d_H;\cO^*_\infty)= H^{<n}(X),
\qquad  H^0(\dl;\cO^*_\infty)=H^n(X), \qquad H^k(\dl;\cO^*_\infty)=0.
\ee
Hence, every $d_H$-closed form $\f\in
\cO^{0,m<n}_\infty$ splits into the sum
\mar{aa3}\beq
\f=\varphi + d_H\xi, \qquad \xi\in \cO^{0,m-1}_\infty, \label{aa3}
\eeq
where $\varphi$ is a closed form on $X$. 

In the case of an affine bundle $Y\to X$, horizontal complexes
(\ref{t70}) -- (\ref{t70a}) induce similar complexes on the base $X$ as
follows. It is quite important for the cohomology calculation of polynomial
complexes in next Section.

Let us consider the open surjection $\pi^\infty:J^\infty
Y\to X$ and the direct image
$\{\pi^\infty_*\gQ^*_\infty\}$ on $X$ of the sheaf $\gQ^*_\infty$. Its stalk
at a point $x\in X$ consists of the equivalence classes of sections of
the sheaf $\gQ^*_\infty$ which coincide on the inverse images
$(\pi^\infty)^{-1}(U_x)$ of neighbourhoods $U_x$ of $x$. Put further the
notation
$\gQ\gX^*_\infty=\pi^\infty_*\gQ^*_\infty$. Since
$\pi^\infty_*\Bbb R=\Bbb R$, we have the following complex of sheaves on $X$:
\mar{t71}\beq
0\to\Bbb R\to \gQ\gX^0_\infty \ar^{d_H}\gQ\gX^{0,1}_\infty\ar^{d_H}\cdots  
\op\longrightarrow^{d_H} 
\gQ\gX^{0,n}_\infty\ar^{d_H} 0.  \label{t71}
\eeq
Every point
$x\in X$ has a base of open contractible neighbourhoods $\{U_x\}$ such that 
the sheaves $\gQ^{0,*}_\infty$ of $\cQ^*_\infty$-modules
are acyclic on the inverse
images 
$(\pi^\infty)^{-1}(U_x)$ of these neighbourhoods. Then, in
accordance with the Leray theorem \cite{god}, cohomology of $J^\infty
Y$ with coefficients in the sheaves $\gQ^{0,*}_\infty$ are isomorphic
to that of $X$ with coefficients in their direct images $\gQ\gX^{0,*}_\infty$,
i.e., the sheaves $\gQ\gX^{0,*}_\infty$ on $X$ are acyclic.
Furthermore, Lemma \ref{am12} shows that
the complexes of 
sections of sheaves $\gQ^{0,<n}_\infty$ over
$(\pi^\infty_0)^{-1}(U_x)$ are exact. It follows that the 
horizontal complex (\ref{t71}), except the last term, 
is also exact.
Due to the
$\Bbb R$-algebra isomorphism
$\cQ^*_\infty=\G(X,\gQ\gX^*_\infty)$, one can think of the
horizontal complex (\ref{t70}) as being the complex of the structure algebras
of sheaves of the horizontal complex (\ref{t71}) on $X$. 

\section{Cohomology of polynomial complexes}

Given the sheaf $\gQ\gX^*_\infty$ on $X$, let us consider its subsheaf
$\gP^*_\infty$ of germs of exterior forms which are polynomials in the fiber
coordinates
$y^i_\La$,
$|\La|\geq 0$, of the topological fiber bundle $J^\infty Y\to X$. This
property is coordinate-independent due to the transition functions
(\ref{55.21}). The $\gP^*_\infty$ is a sheaf of $C^\infty(X)$-modules. Its
structure algebra
$\cP^*_\infty$ is a
$C^\infty(X)$-subalgebra of $\cQ^*_\infty$. For short, one can say that
$\cP^*_\infty$ consists of exterior forms on $J^\infty Y$ which are locally
polynomials in fiber coordinates
$y^i_\La$. 

We have the subcomplex 
\mar{t44}\beq
0\to \Bbb R \ar\gP^0_\infty\ar^{d_H} \gP^{0,1}_\infty\ar^{d_H}\cdots
\ar^{d_H}\gP^{0,n}_\infty \label{t44}
\eeq
of the horizontal complex (\ref{t71})
on $X$. As a particular variant of the algebraic Poincar\'e
lemma, the exactness of the complex (\ref{t44}) has been repeatedly
proved (see, e.g., \cite{barn00}). Since the sheaves  
$\gP^{0,*}_\infty$ of
$C^\infty(X)$-modules on $X$ are acyclic, the
complex (\ref{t44}) is a resolution of the constant sheaf
$\Bbb R$ on $X$. Hence, cohomology of the complex 
\mar{t45}\beq  
0\to \Bbb R \ar\cP^0_\infty\ar^{d_H} \cP^{0,1}_\infty\ar^{d_H}\cdots
\ar^{d_H}\cP^{0,n}_\infty \label{t45}
\eeq
of the structure algebras $\cP^{0,<n}_\infty$ of sheaves $\gP^{0,<n}_\infty$ is
equal to de Rham cohomology of
$X$. It follows that every
$d_H$-closed polynomial form $\f\in\cP^{0,m<n}_\infty$ splits into the sum
\mar{t72}\beq
\f=\varphi + d_H\xi, \qquad \xi\in \cP^{0,m-1}_\infty, \label{t72}
\eeq
where $\varphi$ is a closed form on $X$. 
Let $P^*_\infty$ be $C^\infty(X)$-subalgebra of the polynomial
algebra $\cP^*_\infty$ which 
consists of exterior forms  which are polynomials in the fiber coordinates
$y^i_\La$. Obviously, $P^*_\infty$ is a subalgebra of $\cO^*_\infty$. As a
repetition of Theorem
\ref{am11}, one can show that $P^*_\infty$ have the same cohomology as
$\cP^*_\infty$, i.e., if $\f$ in the decomposition (\ref{t72}) is an
element of
$P^{0,*}_\infty$, then
$\xi$ is so. 

Let us consider the subsheaf $\ol\gP^*_\infty$ of the sheaf 
$\gP^*_\infty$ which consists of germs of $x$-independent polynomial forms.
Its structure algebra $\ol P^*_\infty$ is a subalgebra of the algebra
$P^*_\infty$. We have the complex of sheaves
\be
0\to \Bbb R \ar\ol\gP^0_\infty\ar^{d_H} \ol\gP^{0,1}_\infty\ar^{d_H}\cdots
\ar^{d_H}\ol\gP^{0,n}_\infty 
\ee
which fails to be exact. The obstruction to its exactness at the
term
$\ol\gP^{0,k}_\infty$ is provided by the germs of $k$-forms on $X$ with
constant coefficients \cite{barn00}. Let us denote the sheaf of such germs
by
$S^k_X$.  
For any
$0<k<n$, we have the short exact sequence of sheaves 
\be
0\to \im d_H \to \Ker d_H \to S^k_X \to 0
\ee
and the sequence of their structure modules
\be
0\to \G(X,\im d_H) \to \G(X,\Ker d_H) \to \G(X,S^k_X) \to 0
\ee
which is exact because $S^k_X$ is a subsheaf of $\Bbb R$-modules of the sheaf
$\Ker d_H$. Then, the $k$th cohomology group of the horizontal complex 
\be
0\to \Bbb R \ar\ol P^0_\infty\ar^{d_H} \ol P^{0,1}_\infty\ar^{d_H}\cdots
\ar^{d_H}\ol P^{0,n}_\infty 
\ee
of the algebra $\ol P^*_\infty$ is isomorphic to the $\Bbb R$-module
$\G(X,S^k_X)$ of global constant
$k$-forms on the manifold $X$. 
If a manifold $X$ does not admit an affine coordinate atlas, the module
$\G(X,S^{0,k}_\infty)$  is empty and, consequently, the
differential 
$d_H$ is exact on the algebra $\ol\cP^{0,<n}$. 
Otherwise,  any $d_H$-closed element $\f\in \ol
P^{0,k}_\infty$, $0< k< n$, splits into the sum
$\f=\varphi + d_H\xi$, $\varphi\in \G(X,S^k_X)$, $\xi\in \ol
P^{0,k-1}_\infty$.

\section{Iterated cohomology}

Turn to the proof of Theorem \ref{aa2}. By assumption, the horizontal complex
(\ref{t70'}) is a $\bs|d_H$-bicomplex. 
Its iterated cohomology group $H^{k,m}(\bs|d_H)=E_2^{k,m}$,
$k\in\Bbb Z$, $0\leq m\leq n$, consists of $d_H$-closed horizontal
$m$-forms
$\om\in \cO^{0,m}$ of ghost number $k$ such that $\bs \om$ is $d_H$-exact,
which are taken modulo exterior forms $\bs \psi +d_H\xi$ where $\psi$ is a
$d_H$-closed form.  Then, the  assertions (i) and (ii) of Theorem \ref{aa2}
follows immediately from the decomposition (\ref{aa3}) and the assumption
that forms on $X$ are of zero ghost number and are not
$\bs$-exact. The proof of assertions (iii) -- (v) is based on the analysis of
descent equations.
Since the operator
$\bs$ annihilates exterior forms on
$X$, descent equations can be constructed, but their right-hand side is
not necessarily zero. The key point lies in the existence of closed forms on
$X$ which are exact with respect to the total operator $\wt\bs=\bs +d_H$.

{\it Proof of (iii)}. Let $\om_n$ be a representative of 
the iterated cohomology group $H^{k,n}(\bs|d_H)$, $k<-n$ or $k\geq -1$. It is
a horizontal $n$-form of ghost number $k$
such that
$\bs\om_n$ is
$d_H$-exact, i.e.,
\mar{aa10}\beq
\bs\om_n + d_H\om_{n-1}=0. \label{aa10}
\eeq
Acting  
 on this equality by $\bs$, we observe that $\bs\om_{n-1}$ is a 
$d_H$-closed form of non-vanishing ghost number $k+2$. Therefore, it
is $d_H$-exact, i.e.,
\be
\bs\om_{n-1} + d_H\om_{n-2}=0. 
\ee
Iterating the arguments, one concludes the existence of a family
$\{\om_{n-p}\}$,
$0\leq p\leq n$, of horizontal
$(n-p)$-forms $\om_{n-p}$ of non-vanishing ghost numbers $k+p$ which obey the
descent equations
\mar{aa11}\beq
\bs\om_{n-p} + d_H\om_{n-p-1}=0, \qquad \bs\om_0=0, \qquad 0\leq
p<n.\label{aa11}
\eeq
It may happen that $\om_{n-p}=0$, $p\geq p_0$, for some $p_0$.  Put 
\mar{aa13}\beq
\wt \om_n=\op\sum_{p=0}^n\om_{n-p}. \label{aa13}
\eeq
It is an $\wt\bs$-closed form of total ghost number $k+n$. 
Let
$\{\om_{n-p}'\}$ be another solution of the descent equations (\ref{aa11})
for given $\om_n$. It is readily observed that
$\wt\om_n-\wt\om_n'$ is an
$\wt\bs$-exact form. 
Let the iterated cohomology class of $\om_n$ be zero, i.e.
$\om_n=\bs\xi_n + d_H\xi_{n-1}$ where $\xi_n$ is $d_H$-closed. Then, 
$\{\om_n,\bs\xi_{n-1},0,\cdots,0\}$
is a solution of the descent equations such that
$\wt\om_n$ (\ref{aa13}) 
is an $\wt\bs$-exact form. 
Conversely, let a horizontal exterior form $\wt\om$ of total
ghost number $n+k$ be $\wt\bs$-closed. It splits into the sum (\ref{aa13})
whose summands obey the descent equations (\ref{aa11}). The
higher term $\om_n$ of this sum fulfills the relation (\ref{aa10}), i.e., is
a representative of iterated cohomology. Since all $d_H$-closed
forms of non-vanishing ghost number are $d_H$-exact, the descent
equations (\ref{aa11}) show that: (i) if
$\om_n=0$, then
$\wt\om$ is $\wt\bs$-exact, and (ii) if $\wt\om=\wt\bs\xi$ is $\wt\bs$-exact,
then
$\om_n=\bs\xi_n +d_H\xi_{n-1}$ is of zero cohomology class.
Thus, we come to a desired isomorphism.

{\it Proof of (iv)}. In contrast with the previous case, $d_H$-closed forms
now are not necessarily $d_H$-exact. Therefore, a
representative $\om_n$ of iterated cohomology defines a system of
descent equations where the descent equation of
vanishing ghost number
\mar{aa15}\beq
\bs \om_{n+k+1} +d_H \om_{n+k}=\varphi \label{aa15}
\eeq
has  a closed $(n+k+1)$-form $\varphi$ on $X$ in its right-hand side.
Accordingly, the form
$\wt\om_n$ (\ref{aa13}) fulfills 
 the equality $\wt\bs\wt\om_n=\varphi$.
It follows that the system of descent equations
\mar{aa16}\beq
\wt\bs\wt\om^\varphi=\varphi, \label{aa16}
\eeq
which contains
the equation (\ref{aa15}), admits a solution $\wt\om^\varphi$ iff the
cohomology class of $\varphi$ belongs to $\Ker\g_{n+k+1}$. 
A higher term
$\om_n^\varphi$ of every such solution obeys the
relation (\ref{aa10}) and, consequently, is a representative of iterated
cohomology. If
$\om'^\varphi_n=\om^\varphi_n+ \bs\xi_n +d_H\xi_{n-1}$ is another
representative of the same cohomology class, then $\wt\om+\bs\xi_{n-1}$ is
also a solution of the same descent equations (\ref{aa16}). Consequently,  any
closed $(n+k+1)$-form
$\varphi$ on
$X$ whose cohomology class belongs to $\Ker \g_{n+k+1}$ defines a
subset $A_\varphi$ of the iterated cohomology group $H^{k,n}(\bs|d_H)$, given
by the higher terms of solutions of the descent equations (\ref{aa16}). In
particular, let us consider
$A_{\varphi=0}$. The difference from the proof of item (iii) lies in the
fact that, if the higher term
$\om_n$ of an
$\wt\bs$-closed form $\wt\om$ vanishes, then $\wt\om=\wt\bs\xi +\psi$ where
$\psi$ is a $(n+k)$-closed form on $X$. It follows that $A_{\varphi=0}=\ol
H^{k+n}$. One can justify easily that: (i) $\om_n^\varphi\in A_\varphi$ and
$\om^0_n\in A_{\varphi=0}$ implies $\om_n^\varphi+\om^0_n\in
A_\varphi$, and (ii)
$\om^\varphi_n, \om'^\varphi_n\in A_\varphi$ implies
$\om'^\varphi_n-\om^\varphi_n\in A_{\varphi=0}$. It follows that $A_\varphi$
is an affine space modelled over the linear space  $\ol H^{k+n}$. Let
$\varphi'=\varphi+d\si$. Then, a solution $\wt\om^\varphi$ of the descent
equations (\ref{aa16}) defines  a solution
$\wt\om^\varphi +\si$ of the descent equations (\ref{aa16}) where $\varphi$ is
replaced with $\varphi'$. These solutions have the same higher term
$\om^\varphi_n=\om_n^{\varphi'}$ and, consequently define the same
representative of iterated cohomology. It follows that
$A_\varphi=A_{\varphi+d_H\si}$, i.e., $A_\varphi$ is set by the cohomology
class
$[\varphi]\in \Ker\g_{k+n+1}$ of $\varphi$. It remains
to show that $H^{k,n}(\bs|d_H)$ is a disjoint union of sets
$A_{[\varphi]}$,
$[\varphi]\in \Ker\g_{k+n+1}$. Indeed, let a representative of
iterated cohomology defines different systems of descent equations
(\ref{aa16}) with
$\varphi$ and
$\varphi'$ in the right-hand side. Then, one can easily justify that 
$\varphi'-\varphi$ is an
exact form. Turn to the particular case $k=-n$. If a constant function on $X$ 
is  $\wt\bs$-exact, it is
$\bs$-exact. Therefore,  
$\Ker\g=0$.

{\it Proof of (v)}. In this case, we have the descent equations (\ref{aa11})
with the zero right-hand side, but $H^{-1,n}(\bs|d_H)=H^{-1}_{\rm tot}/\im
\g_{n-1}$. 

Note that, in the case of BRST cohomology (see Corollary \ref{aaa}), the
right-hand side of the global descent equations for any total ghost number
remains zero.


\begin{thebibliography}{bbb}

\bibitem{ander80} Anderson, I. and Duchamp, T.: {\it Amer. J. Math.} {\bf 102}
(1980) 781.

\bibitem{ander} Anderson, I.: {\it
Contemp. Math.} {\bf 132} (1992) 51.

\bibitem {barn}  Barnish, G., Brandt, F. and Henneaux, M.: {\it Commun. Math.
Phys.} {\bf 174} (1995) 57.

\bibitem{barn00} Barnish, G., Brandt, F. and Henneaux, M.: {\it Phys. Rep.}
(appear); e-print: hep-th/0002245.

\bibitem{bau} Bauderon, M.: In: {\it Differential Geometry, Calculus of
Variations, and their Applications}, Lecture Notes in Pure and Applied
Mathematics, 100, Marcel Dekker Inc., N.Y., 1985, pp. 67-82.

\bibitem{brandt}  Brandt, F.: {\it Commun.
Math. Phys.} {\bf 190}  (1997) 459.

\bibitem{bred}  Bredon, G.: {\it Sheaf Theory},
McGraw-Hill Book Company, New York, 1967. 

\bibitem{bred2} Bredon, G.: {\it Topology and Geometry},
Graduate Texts in Mathematics, 139, Springer-Verlag, Berlin, 1997. 

\bibitem{db92} Duboes--Violette, M., Henneaux, M., Talon, M. and
Viallet, C.-M.: {\it Phys. Lett.} {\bf B289} 361.

\bibitem{book} Giachetta, G., Mangiarotti, L. and Sardanashvily, G.: {\it New
Lagrangian and Hamiltonian Methods in Field Theory}, World Scientific,
Singapore, 1997.

\bibitem{eprint} Giachetta, G., Mangiarotti L. and Sardanashvily, G.:
 e-print: math-ph/0005010. 

\bibitem{god}  Godement, R.: {\it Th\'eorie des faisceaux}, Hermann,
Paris, 1964. 

\bibitem{henn91} Henneaux, M.: 
{\it Commun. Math. Phys.} {\bf 140} (1991) 1.

\bibitem{hir} Hirzebruch, F.: {\it Topological Methods in Algebraic Geometry},
Springer-Verlag, Berlin, 1966.

\bibitem{kru98} Krupka, D. and Musilova, J.: {\it Diff. Geom. Appl.} {\bf
9} (1998) 293.

\bibitem{mcl} Mac Lane, S.: {\it Homology}, Springer-Verlag, Berlin, 1967.

\bibitem{book00} Mangiarotti, L. and Sardanashvily, G.: {\it Connections in
Classical and Quantum Field Theory}, World Scientific, Singapore, 2000.

\bibitem{olver} Olver, P.: {\it Applications of Lie Groups to
Differential Equations}, Springer-Verlag, Berlin, 1997. 

\bibitem{tak2} Takens, F.:{\it J. Diff. Geom.} {\bf 14} (1979) 543. 

\bibitem{tul} Tulczyjew, W.: 
In: {\it Differential
Geometric Methods in Mathematical Physics}, Lect. Notes in Mathematics,
836, Springer-Verlag, Berlin, 1980, pp. 22-48.

\bibitem{vin} Vinogradov, A.: {\it J. Math. Anal.
Appl.} {\bf 100} (1984) 41.

\end{thebibliography}
\end{document}